\documentclass[aps,prb,twocolumn,groupedaddress,floatfix]{revtex4}

\usepackage{graphicx}
\usepackage{subfigure}

\def\be{\begin{equation}}
\def\ee{\end{equation}}
\def\bea{\begin{eqnarray}}
\def\eea{\end{eqnarray}}

\begin{document}


\title{Electronic instabilities in 3D arrays of small-diameter 
(3,3) carbon nanotubes}


\author{J. Gonz{\'a}lez$^{1}$ and E. Perfetto$^{1,2}$}
\affiliation{$^{1}$Instituto de Estructura de la Materia.
        Consejo Superior de Investigaciones Cient{\'\i}ficas.
        Serrano 123, 28006 Madrid. Spain.\\
$^{2}$Istituto Nazionale di Fisica Nucleare - Laboratori
Nazionali di Frascati, Via E. Fermi 40, 00044 Frascati, Italy.}


\date{\today}

\begin{abstract}
We investigate the electronic instabilities of the small-diameter
(3,3) carbon nanotubes by studying the low-energy perturbations of the 
normal Luttinger liquid regime. The bosonization approach is adopted
to deal exactly with the interactions in the forward-scattering
channels, while renormalization group methods are used to analyze
the low-energy instabilities. In this respect, we take into account
the competition between the effective e-e interaction mediated by
phonons and the Coulomb interaction in backscattering and Umklapp
channels. Moreover, we apply our analysis to relevant experimental
conditions where the nanotubes are assembled into large 
three-dimensional arrays, which leads to an efficient screening of 
the Coulomb potential at small momentum-transfer. We find that the 
destabilization of the normal metallic behavior takes place through 
the onset of critical behavior in some of the two charge stiffnesses 
that characterize the Luttinger liquid state. From a physical point 
of view, this results in either a divergent compressibility or a 
vanishing renormalized velocity for current excitations at the point 
of the transition. We observe anyhow that this kind of critical 
behavior occurs without the development of any appreciable sign 
of superconducting correlations.

\end{abstract}

\maketitle

\vspace*{1cm}

The development of nanoscale technology during the last decade 
has attracted much attention on carbon nanotubes, which are
among the most promising candidates to fabricate molecular-size
devices. This is mainly due to the wide variety of their electronic 
and transport properties, which can result in metallic\cite{mint}, 
semiconducting\cite{saito} or even superconducting behavior\cite{kas}, 
depending on geometry and the way of assembling.

>From a theoretical point of view, the confinement of electrons in
the longitudinal dimension of the nanotubes induces the so-called 
Luttinger liquid behavior\cite{bal,eg,kane,yo,tse}. This is 
characterized, for instance, by the power-law dependence of the 
differential conductance, which has been actually observed 
experimentally\cite{exp,yao}.
 
Such a behavior breaks down anyhow at sufficiently low temperature, and 
the nanotubes enter a different regime, usually driven by the quality 
of the contacts in the experimental setup. In particular, in the case of
very transparent contacts, it has been observed that carbon nanotubes
may develop superconducting correlations, inherited from superconducting
electrodes (proximity effect)\cite{kas} as well as intrinsic to large 
assemblies of massive ropes\cite{sup,priv}.

The superconducting instability is anyway in competition with the
so-called Peierls (or charge-density-wave) instability, which may
induce a metallic-semiconducting transition caused by a lattice
distorsion. The mean-field temperature $T_P$ of such a transition has
been estimated by means of detailed calculations, and it is predicted
to increase as the radius of the nanotubes becomes smaller\cite{peierls}. 
For tubes of typical radius, calculations find a very 
low (undetectable) value of $T_P$, while for thinner nanotubes it is
predicted to be significantly larger and competing with the 
superconducting critical temperature. 

Nevertheless, superconductivity at about 15 K
has been claimed to occur in $4$ {\AA}-diameter nanotubes\cite{tang}.
In the experiment reported in Ref. \onlinecite{tang}, a strong
diamagnetic behavior was interpreted as an anisotropic Meissner
effect, while a genuine superconducting transition was not observed. 
This has opened some controversy on this issue, since {\it ab initio}
simulations predict a room-temperature Peierls transition in the
allowed $4${\AA}-diameter geometries, namely in the (5,0)\cite{ab1} 
and the (3,3) nanotubes\cite{ab2}. On the other hand, mean-field 
calculations for the (5,0) nanotubes seem to find a superconducting 
instability, but with a critical temperature of about 1 K\cite{mf}.

In this paper we investigate the low-energy properties of the (3,3)
nanotubes by focusing on the instabilities of the Luttinger liquid 
behavior. We study carefully the competition between the effective 
e-e interaction mediated by phonons and the Coulomb repulsion. The 
bosonization technique is applied in order to deal exactly with the 
interactions in the forward-scattering channels, while renormalization 
group methods are used to approach the low-energy instabilities of 
the system, driven by the backscattering and Umklapp interactions.  
Moreover, we pay also special attention to the experimental conditions 
reported in Ref. \onlinecite{tang}, which lead to large arrays of
nanotubes embedded in a zeolite matrix. This gives rise to a large 
screening of the Coulomb potential, which has no counterpart in the 
case of single nanotubes\cite{th1,th2}. We study this effect by means 
of a generalized RPA approach, showing that the long-range intertube 
coupling produces an efficient screening of the intratube interactions 
with small momentum-transfer.

The most important result of the present study is the finding of two
different low-energy phases characterized by critical (nonanalytic)
behavior of the physical observables. Under the conditions corresponding 
to the experimental samples described in Ref. \onlinecite{tang}, the 
critical behavior is related to the vanishing of one of the Luttinger
liquid parameters, and it is qualitatively consistent with the large
diamagnetic signal observed in Ref. \onlinecite{tang}. We also observe
that this kind of singularity occurs well before the development of
any sizeable superconducting or charge-density-wave correlations
in the electron system.


We start by paying attention to the interactions mediated by the 
Coulomb potential, which provides a strong source of repulsion between
electrons in single nanotubes. In the tubular nanotube geometry, the 
Coulomb potential is given by\cite{eg} 
\begin{equation}
V_C ({\bf r}-{\bf r}')=\frac{ e^{2} /\kappa }
 {\sqrt{(x-x')^{2}+4R^{2}\sin^{2}[(y-y')^{2}/2R^{2}]+a_{z}^{2} }}
\label{potent}
\end{equation}
where $a_{z} \simeq 1.6 \; {\rm \AA}$  and $R$ is the nanotube radius.
The screening by the environment of external charges is in general
encoded in the dielectric constant $\kappa $. For the sake of 
studying the nanotube transport properties, it is more convenient
do deal with the one-dimensional
(1D) projection of the potential onto the longitudinal
dimension of the nanotube. This is achieved by integration of the 
circular coordinate, upon which we obtain the 1D Coulomb potential
$\tilde{V}_C (k)$ depending on the longitudinal momentum-transfer 
$k$ \cite{sar}
\begin{equation}
 \tilde{V}_C (k) \approx
         \frac{2 e^2 }{\kappa }  \log(\frac{k_c+k}{k}) \, ,
\label{Coulomb}
\end{equation}
$k_c$ is in general of the order of the inverse of the
nanotube radius $R$, as it is the memory that the 1D projection
keeps of the finite transverse size. 

\begin{figure}
\includegraphics[width =7.5cm ,height=7.5cm ]{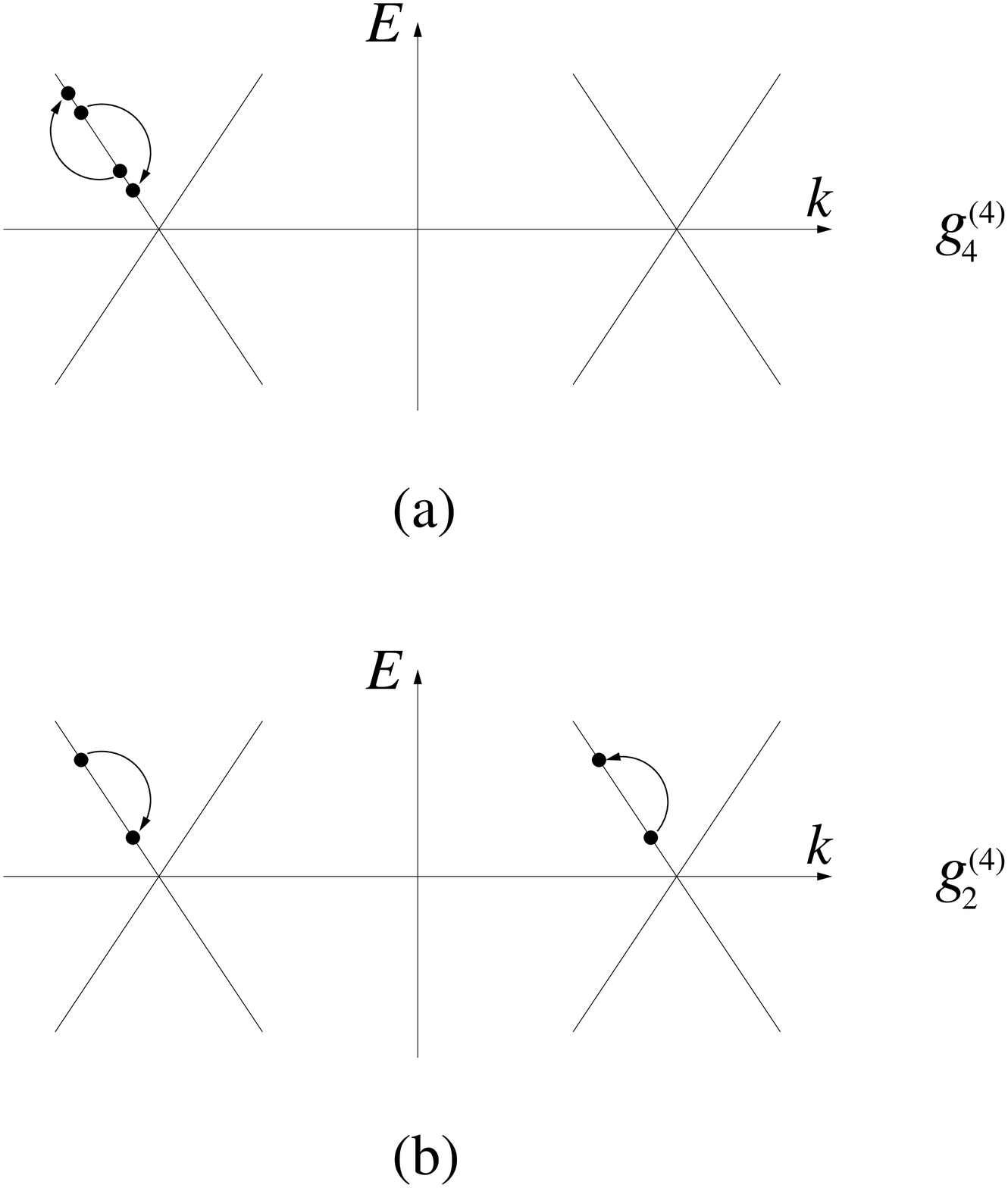}
\caption{Small momentum-transfer processes corresponding
to the couplings $g_4^{(4)}$ and $g_2^{(4)}$.}
\label{g4424}
\end{figure}

\begin{figure}
\includegraphics[width =7.5cm ,height=7.5cm ]{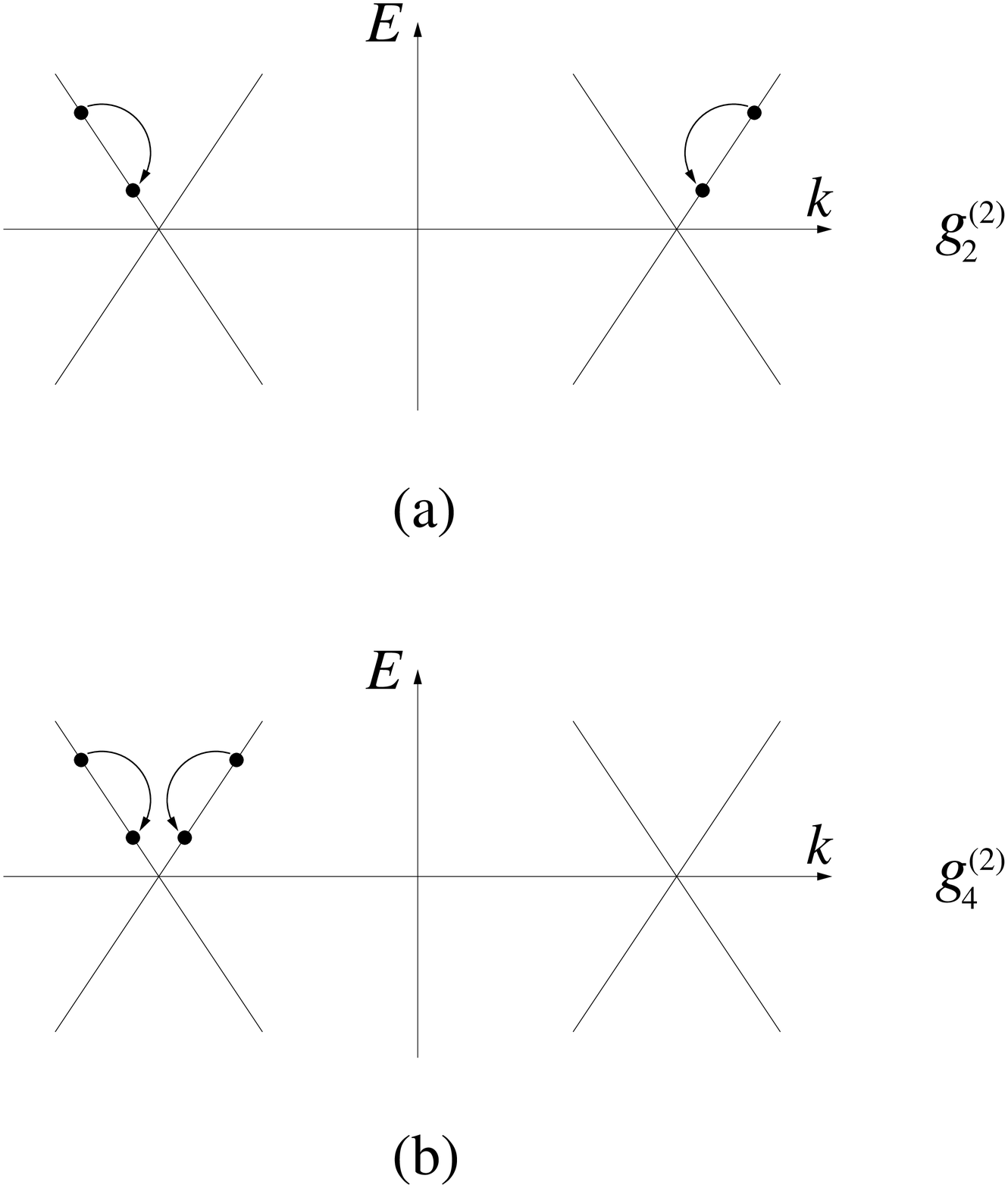}
\caption{Small momentum-transfer processes corresponding
to the couplings $g_2^{(2)}$ and $g_4^{(2)}$.}
\label{g2242}
\end{figure}

The relative strength of the Coulomb interaction is given by the 
dimensionless ratio between $e^2$ and the Fermi velocity $v_F$, 
which for the (3,3) nanotubes is approximately $e^2 /v_F \sim 2.9$. 
This means that the Coloumb potential should give the dominant 
interaction in the forward-scattering channels, at least in single 
nanotubes. The processes can be classified depending on the 
low-energy linear branches involved in the electron-electron 
scattering\cite{louie}. We recall at this point that the low-energy 
modes of the (3,3) nanotubes lie in a bonding and an antibonding 
subband that cross at two Fermi points (in the undoped system) with 
opposite longitudinal momenta\cite{disper}.
We can distinguish in particular four forward-scattering channels, 
labelled by their respective couplings as represented in Figs. 
\ref{g4424} and \ref{g2242}. In these processes there is a nominally 
strong Coulomb repulsion between the electrons, as they scatter 
without change of their chirality and the interaction strength
is simply given by the Coulomb potential (\ref{Coulomb}).

The interaction channels represented in Figs. \ref{g4424} and
\ref{g2242} are only part of the complete catalogue of scattering
processes, that may be classified according to the chiralities and
momentum-transfer for the interacting electrons\cite{louie}. 
It has become 
usual to assign respective coupling constants $g_i^{(j)}$ to the 
interaction channels, in such a way that the lower index discerns 
whether the interacting particles shift from one Fermi point to the 
other $(i=1)$, remain at different Fermi points $(i=2)$, or they
interact near the same Fermi point $(i=4)$. The upper label
follows the same rule to classify the different combinations of
left-movers and right-movers, including the possibility of
having Umklapp processes $(j=3)$. As we will see, the couplings
for the channels with large momentum-transfer $2k_F$ or change of 
chirality of the interacting electrons have an strength which is 
sensibly smaller than that of the forward-scattering couplings
represented in Figs. \ref{g4424} and \ref{g2242}.

\begin{figure}
\includegraphics[width =7.5cm ]{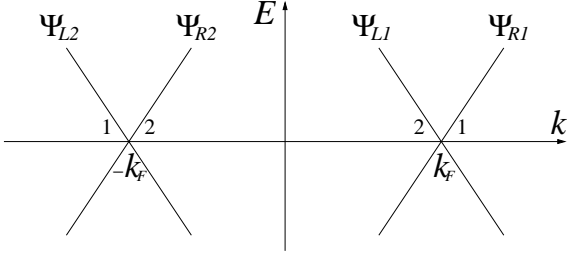}
\caption{Schematic representation of the low-energy branches arising 
from the bonding and the antibonding subband of the armchair nanotubes.}
\label{bran}
\end{figure}

A very important point is that the system with just the 
forward-scattering interactions $g_4^{(4)}, g_2^{(4)}, g_2^{(2)}$
and $g_4^{(2)}$ is exactly solvable by means of bosonization 
techniques\cite{emery,sol,gog}. 
Thus, no matter that the Coulomb repulsion may place
these interaction channels in the strong-coupling regime, the 
availability of an exact solution makes possible to capture 
the nonperturbative effects coming from the Coulomb interaction.
The bosonization techniques make use of the fact that the 
forward-scattering interactions can be written in terms of
the electron density operators
corresponding to the different electron fields
$\Psi_{r i \sigma } $ for the linear branches shown in Fig. 
\ref{bran}. We adopt a notation in which the index
$r = L, R$ is used to label the left- or right-moving character
of the linear branch, and the index $i = 1,2 $ to label the Fermi
point. The index $\sigma $ stands for the two different spin
projections. We may actually introduce the charge and spin
density operators
\begin{eqnarray}
   \rho_{r i  }(x)  & = &
 \frac{1}{\sqrt{2}}
    ( \Psi^{\dagger}_{r i \uparrow }(x) \Psi_{r i \uparrow }(x) +
 \Psi^{\dagger}_{r i \downarrow }(x) \Psi_{r i \downarrow }(x) ) \\
   \sigma_{r i  }(x)  & = &
 \frac{1}{\sqrt{2}}
 (   \Psi^{\dagger}_{r i \uparrow }(x) \Psi_{r i \uparrow }(x) -
 \Psi^{\dagger}_{r i \downarrow }(x) \Psi_{r i \downarrow }(x)  )
\end{eqnarray}
As long as the Coulomb interaction and the interaction mediated by 
exchange of phonons do not depend on the spin of the interacting
electrons, we will carry out the subsequent discussion in terms 
of the charge density operators $\rho_{r i  }(x)$.

It is convenient, for instance, to introduce operators for the 
sum and the difference of charge densities in the bonding and the 
antibonding subbands of the armchair nanotubes:
\begin{eqnarray}
\rho_{R \pm  }(k) & = &   \frac{1}{\sqrt{2}}
          (  \rho_{R 1  }(k) \pm  \rho_{R 2  }(k) )     \\
\rho_{L \pm  }(k) & = &   \frac{1}{\sqrt{2}}
          (  \rho_{L 2  }(k) \pm  \rho_{L 1  }(k)  )   \, .
\end{eqnarray}
In terms of these operators, the hamiltonian for the
forward-scattering interactions can be written in the form
\begin{eqnarray}
H_{FS}  & = &    \frac{1}{2} v_F \int_{-k_c}^{k_c} dk
 \sum_{r=L,R} \sum_{i=\pm }   \rho_{r i } (k)
             \rho_{r i } (-k)     \nonumber    \\
  &  &    + \frac{1}{2}  \int_{-k_c}^{k_c} \frac{dk}{2\pi } \;
    2   \left(   \rho_{R +} (k) \;
       (g_4^{(4)} + g_2^{(4)})     \;
   \rho_{R +} (-k)   \right.     \nonumber     \\
  &   &    +    \rho_{L +} (k) \;
       (g_4^{(4)} + g_2^{(4)})     \;
   \rho_{L +} (-k)           \nonumber     \\
  &  &    +   \rho_{R -} (k) \;
       (g_4^{(4)} - g_2^{(4)})     \;
   \rho_{R -} (-k)        \nonumber     \\
  &   &    +    \rho_{L -} (k) \;
       (g_4^{(4)} - g_2^{(4)})     \;
   \rho_{L -} (-k)           \nonumber     \\
  &  &    +  2  \rho_{R +} (k) \;
       (g_2^{(2)} + g_4^{(2)})     \;
   \rho_{L +} (-k)       \nonumber     \\
  &   &  \left.  +  2  \rho_{R -} (k) \;
       (g_2^{(2)} - g_4^{(2)})     \;
   \rho_{L -} (-k)   \right)
\label{ham2}
\end{eqnarray}
where $k_c$ stands again for the momentum cutoff 
dictated by the transverse size of the nanotube.

We observe that the symmetric and the antisymmetric combination
of the charge operators in the two low-energy subbands decouple 
in the hamiltonian (\ref{ham2}). This can be completely diagonalized
by first introducing boson fields $\Phi_+ (x)$ and $\Phi_- (x)$ (and 
their respective conjugate momenta, $\Pi_+ (x)$ and $\Pi_- (x)$) 
defined by 
\begin{eqnarray}
\partial_x \Phi_+ (x)  & = &  \sqrt{\pi }
           (  \rho_{L +}(x) + \rho_{R + }(x)  )        \\
\partial_x \Phi_- (x)  & = &  \sqrt{\pi }
           (  \rho_{L -}(x) + \rho_{R - }(x)  )
\end{eqnarray}
The hamiltoninan can be rewritten then in the form
\begin{eqnarray}
H_{FS}   & = &   \frac{1}{2} u_+ \int dx
   \left(  K_+ ( \Pi_{+} (x)  )^2  +
 \frac{1}{K_+} ( \partial_x \Phi_{+}(x) )^2  \right)    \nonumber  \\
   &   &  +  \frac{1}{2} u_- \int dx
    \left(  K_- ( \Pi_{-} (x)  )^2  +
 \frac{1}{K_-} ( \partial_x \Phi_{-}(x) )^2  \right)
\label{hlutt}
\end{eqnarray}
with renormalized velocities $u_+, u_-$ and charge stiffnesses 
$K_+, K_-$ given by the expressions
\begin{eqnarray}
 u_{\pm} K_{\pm}  & = &
        v_F + \frac{1}{\pi }  \left( g_4^{(4)} \pm g_2^{(4)}
                - (g_2^{(2)} \pm g_4^{(2)}) \right)
                                             \label{v2}      \\
 u_{\pm} / K_{\pm}  & = &
     v_F +  \frac{1}{\pi }   \left( g_4^{(4)} \pm g_2^{(4)}
                + (g_2^{(2)} \pm g_4^{(2)}) \right)
                                              \label{v1}
\end{eqnarray}

Upon application of the canonical transformation
\begin{eqnarray}
 \Phi_{+} =  \sqrt{K_+} \tilde{\Phi}_+
\,  &,&  \,
  \Pi_{+} = \frac{1} {\sqrt{K_+} }\tilde{\Pi}_+      \nonumber \\
 \Phi_{-} =  \sqrt{K_-} \tilde{\Phi}_-
\,  &,&  \,
  \Pi_{-} = \frac{1} {\sqrt{K_-} } \tilde{\Pi}_-
\end{eqnarray}
the hamiltonian (\ref{hlutt}) becomes that of a couple of 
noninteracting boson fields. This picture, in which the low-energy 
excitations are just given by charge fluctuations (and spin 
fluctuations with unrenormalized velocity), is what characterizes 
the Luttinger liquid regime of the electron system\cite{emery,sol}.
The charge stiffnesses and the renormalized velocities in the two 
different charge sectors dictate for instance the thermodynamic and
transport properties, encoded into the compressibilities  
$\kappa_{\pm}$, the Drude weights $D_{\pm}$ and the dependence of 
the specific heat $C_{v}$ on the temperature $T$\cite{sch}:
\begin{eqnarray}
\kappa_{\pm}  & = & \frac{2}{\pi} \frac{K_{\pm}}{u_{\pm}}  \label{compr}\\
D_{\pm}  & = &    2 u_{\pm} K_{\pm}    \label{drude}      \\
\frac{C_{v}}{T} & = & \frac{\pi}{3}(\frac{1}{u_{+}}+\frac{1}{u_{-}})
\end{eqnarray}

The above picture will be complemented later by including the effect 
of the backscattering and Umklapp interactions, that tend to destabilize 
the Luttinger liquid regime at low temperatures. At this point, we stress 
that the robustness of such a regime depends on the strength of the 
Coulomb interaction. The consideration of the screening effects induced 
by the environment becomes then quite relevant, specially in the case of 
a large assembly of nanotubes\cite{th1,th2}. For the nanotube array 
described in Ref. \onlinecite{tang}, the long-range Coulomb repulsion 
gives rise to a nonnegligible interaction between electronic currents in
different nanotubes. If we label these by their respective positions
${\bf l}$ and ${\bf l'}$ in the transverse section of the array, 
the intertube Coulomb potential can be expressed as 
\begin{equation}
V_{{\bf l}, {\bf l'}} (k) \approx
\frac{2 e^2}{\kappa } K_0 (| {\bf l}-{\bf l'}| k) \, ,
\label{intert}
\end{equation}
where $k$ denotes the longitudinal momentum-transfer. We recall that
the Bessel function $K_0 (x)$ is logarithmically divergent in the 
limit $x \rightarrow 0$. For ${\bf l} = {\bf l'}$, there is implicit 
a short-distance cutoff given by the radius of the nanotube, which 
leads then to the potential (\ref{Coulomb}). The important point is
that the intertube Coulomb potential gives rise to quite significant 
screening effects at small momentum-transfer, which modify appreciably
the strength of the forward-scattering couplings described above.

\begin{figure}
\includegraphics[width =9cm ]{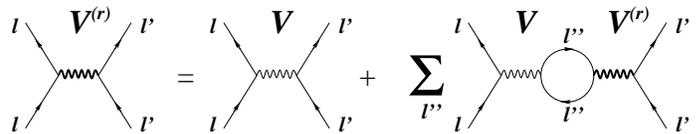}
\caption{Self-consistent diagrammatic equation representing the 
screening of the Coulomb potential between charges at nanotubes
${\bf l}$ and ${\bf l'}$. The sum over ${\bf l''}$ accounts for 
the polarization effects from all the nanotubes in the array.}
\label{self}
\end{figure}

In order to take into account the interaction among all the nanotubes
in the array, we can adopt a generalization of the RPA scheme used in 
Ref. \onlinecite{hawrylak} for the study of 2D layered sytems. In our 
case, the screened Coulomb potential $V_{{\bf l}, {\bf l'}}^{(r)} (k)$
has to satisfy the self-consistent diagrammatic equation shown in 
Fig. \ref{self}, with ${\bf l''}$ running over all the positions of the
nanotubes in the array. We have then 
\begin{equation}
V_{{\bf l},{\bf l'}}^{(r)}(k) = V_{{\bf l},{\bf l'}}(k)
   +\Pi \,  \sum_{{\bf l''}}V_{{\bf  l},{\bf l''}}(k) \,
 V_{{\bf l''},{\bf l'}}^{(r)}(k),
\label{dyson}
\end{equation}
where $\Pi $ stands for the polarization of each 1D electron system.
This function is known exactly at small momentum-transfer\cite{sol}, 
and here it appears multiplied by the number of subbands $n = 2$ 
contributing at low-energies:
\begin{equation}
\Pi (k, \omega) = 2n \frac{1}{\pi } \frac{v_F k^2}{\omega^2 - v_F^2 k^2}
\end{equation}

Equation (\ref{dyson}) can be easily solved by introducing the 
Fourier transform of the Coulomb potential with respect to the 
nanotube position ${\bf l}$ in the 2D transverse section of the array. 
We define for instance $\phi ({\bf p}, k)$ by
\begin{equation}
V_{{\bf l},{\bf l'}} (k) =  
     \left( \frac{d}{2 \pi} \right)^{2} \int_{BZ}  d^{2}{\bf p} \, 
\phi ({\bf p}, k) \,  e^{i {\bf p} \cdot ({\bf l}-{\bf l'})},
\label{fourier}
\end{equation}
where $d$ stands for the nanotube separation and $BZ$ denotes that the 
integration is over the Brillouin zone for the nanotube lattice in the 
transverse section of the array. We define similarly 
$\phi^{(r)} ({\bf p}, k)$ as the Fourier transform of 
$V_{{\bf l},{\bf l'}}^{(r)} (k)$. The expression of equation 
(\ref{dyson}) becomes then in momentum space
\begin{equation}
\phi^{(r)} ({\bf p}, k) = \phi ({\bf p}, k)  +  
   \Pi \, \phi ({\bf p}, k)\, \phi^{(r)} ({\bf p}, k).
\label{algdys}
\end{equation}
In what follows, we will stick to the solution of this equation at
$\omega = 0$, for the sake of giving a simpler description of the 
screening effects in the static limit.

\begin{figure}
\includegraphics[width =7.5cm ]{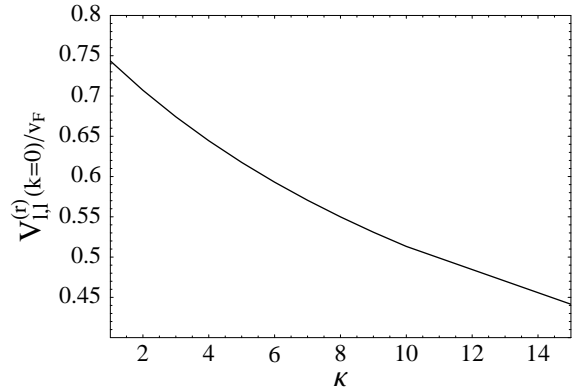}
\caption{Dependence of the screened intratube Coulomb potential 
(in units of $v_F$) on the dielectric constant $\kappa $.}
\label{intra-k}
\end{figure}

Within our RPA scheme, the screened Coulomb potential becomes finally 
\begin{equation}
V_{{\bf l},{\bf l'}}^{(r)}(k)  =  
   \left( \frac{d}{2 \pi} \right)^{2}\int_{BZ} d^{2}{\bf p} 
   \frac{ \phi ({\bf p}, k) }{ 1 + 2n \phi ({\bf p}, k)/\pi v_F }
 e^{i {\bf p}\cdot ({\bf l}-{\bf l'})}.
\label{screened}
\end{equation}
The most important property of $V_{{\bf l},{\bf l'}}^{(r)}(k)$ is 
that it saturates at a finite value in the limit $k \rightarrow 0$.
This is a reflection of the fact that, for distances much larger 
than the nanotube separation, the array screens effectively the 
Coulomb interaction as a 3D system. Thus, taking a value of the 
bare coupling $e^2 \approx 2.9 v_F$, we find that the intratube 
potential at vanishing momentum-transfer is 
$V_{{\bf l},{\bf l}}^{(r)}(k=0) \approx 0.75 v_F$ (for $\kappa = 1$).
This value depends slightly on the dielectric constant, as 
represented in Fig. \ref{intra-k}. We will take this strength of 
the screened intratube Coulomb potential at $ k = 0$
as an input for the values of the forward-scattering couplings 
$g_4^{(4)}, g_2^{(4)}, g_2^{(2)}$ and $g_4^{(2)}$ within each 
nanotube in the array.

\begin{figure}
\includegraphics[width =7.5cm ]{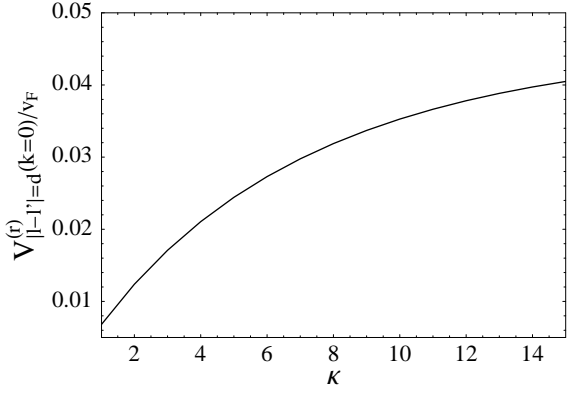}
\caption{Dependence of the screened intertube Coulomb potential
(in units of $v_F$) on the dielectric constant $\kappa $.}
\label{inter-k}
\end{figure}

\begin{figure}
\includegraphics[width =7.5cm ]{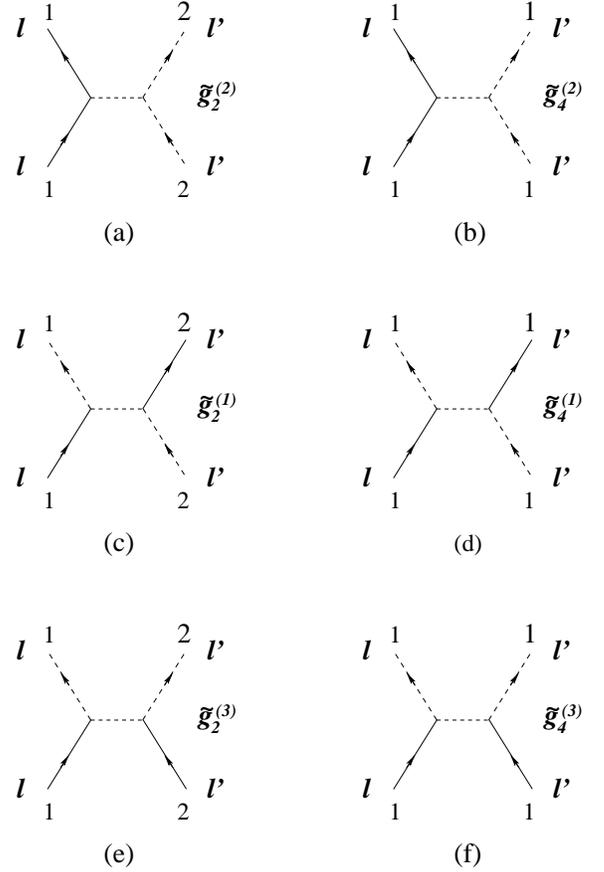}
\caption{Intertube interactions arising from the coupling between
electronic currents in nearest-neighbor nanotubes ${\bf l}$ and
${\bf l'}$ of a 3D array. The full (dashed) lines represent the 
propagation of electrons with right (left) chirality, and the labels
1, 2 denote the respective Fermi points. The dashed lines 
(without arrow) stand for the intertube Coulomb potential.}
\label{gtilde}
\end{figure}

A closer look at the above analysis shows that the Coulomb 
potential between nearest-neighbor nanotubes also gives rise to 
relevant interaction channels. The new couplings needed for a
consistent description of the interaction processes are
catalogued in Fig. \ref{gtilde}. They adhere to the same rules
used to label the intratube couplings, but with a tilde to
distinguish their intertube character. In particular, the 
intertube forward-scattering interactions $\tilde{g}_2^{(2)}$
and $\tilde{g}_4^{(2)}$ (preserving the chirality
of the electrons within each nanotube) are affected by the screening
effects described above, as they are corrected by the polarization
of the different nanotubes in the array. Their strength is given
by the intertube Coulomb potential, which keeps only a nonnegligible 
value between nearest-neighbor tubes, after screening by the 
nanotube array. Taking again $e^2 \approx 2.9 v_F$, the intertube 
potential for $|{\bf l} - {\bf l'}| = d$ has a value at vanishing 
momentum-transfer $V_{{\bf l},{\bf l'}}^{(r)}(k=0) \approx 0.007 v_F$ 
(for $\kappa = 1$). The dependence of this strength on the 
dielectric constant is shown in Fig. \ref{inter-k}. Although the 
relative strength of the intertube forward-scattering couplings 
$\tilde{g}_2^{(2)}$ and $\tilde{g}_4^{(2)}$ may appear small, 
these are however significant as they influence the scaling 
of the rest of interactions at low energies, as we will see in what 
follows.

\begin{figure}
\includegraphics[width =4cm ]{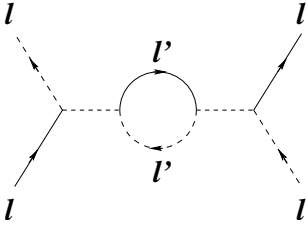}
\caption{Second-order process renormalizing backscattering interactions 
at a given nanotube ${\bf l}$ through the coupling with the 
nearest-neighbors ${\bf l'}$ in a 3D array of nanotubes. The dashed lines 
(without arrow) stand for the intertube Coulomb potential.}
\label{pol}
\end{figure}

On the other hand, the intertube chirality-breaking interactions
$\tilde{g}_2^{(1)}, \tilde{g}_4^{(1)}, \tilde{g}_2^{(3)}$ and 
$\tilde{g}_4^{(3)}$
give rise to a different kind of screening processes, of the type 
represented in Fig. \ref{pol}. The effect of these processes 
cannot be captured in the RPA scheme described above, as the
polarization with a change of chirality in the particle-hole
pair diverges logarithmically at low energies.
As long as the corrections depend on the 1D cutoff, 
they give rise instead to new screening contributions to the 
scaling equations for the rest of interactions. We recall that
the scaling equations for the interactions of single armchair 
nanotubes have been obtained in Ref. \onlinecite{louie} (up to
terms quadratic in the couplings). We have found that, after 
incorporating the corrections from the intertube interactions, 
the scaling equations for the intratube couplings become 
\begin{eqnarray}
\frac{\partial g_1^{(1)}}{\partial l}  & = &
    - \frac{1}{\pi v_F}  (  g_1^{(1)} g_1^{(1)}
        +  g_1^{(2)} g_2^{(1)}               \nonumber     \\
   &  &   + g_1^{(3)} g_1^{(3)} - g_1^{(3)} g_2^{(3)}  ) 
                                               \label{first}   \\
\frac{\partial g_1^{(2)}}{\partial l}  & = &
     (1 - \frac{1}{K_{-}}) g_1^{(2)}
     -   \frac{1}{\pi v_F}  (  g_2^{(1)} g_1^{(1)}  \nonumber  \\
     &  &    - g_4^{(3)} g_1^{(3)}  )                        \\
\frac{\partial g_1^{(3)}}{\partial l}  & = &
        (1 - K_{+})  g_1^{(3)}
   - \frac{1}{\pi v_F}   (   2 g_1^{(3)} g_1^{(1)}   \nonumber \\
  &  &   -  g_2^{(3)} g_1^{(1)}  - g_4^{(3)} g_1^{(2)}  )     \\
\frac{\partial g_2^{(1)}}{\partial l}  & = &
     (1 - \frac{1}{K_{-}})  g_2^{(1)}
   - \frac{1}{\pi v_F}  ( 2  g_4^{(1)} g_2^{(1)} 
     -  g_4^{(1)} g_1^{(2)}    \nonumber     \\
   &  &  +  g_1^{(2)} g_1^{(1)}  + g_4^{(3)} g_2^{(3)}
             - g_4^{(3)} g_1^{(3)}                 \nonumber    \\
  &   &     + 12 \tilde{g}_4^{(1)}   \tilde{g}_2^{(1)}   
            + 12 \tilde{g}_4^{(3)}   \tilde{g}_2^{(3)}      )    \\
\frac{ \partial g_2^{(2)}}{\partial l}  & = &
       -  \frac{1}{2 \pi v_F}  (  g_2^{(1)} g_2^{(1)}
           +  g_1^{(1)} g_1^{(1)}                  \nonumber     \\  
    &  &   +  g_1^{(2)} g_1^{(2)} - g_2^{(3)} g_2^{(3)} )     \\
\frac{ \partial g_2^{(3)}}{\partial l}  & = &
      (1 - K_{+}) g_2^{(3)}
  - \frac{1}{\pi v_F}  ( 2  g_4^{(1)} g_2^{(3)}   \nonumber     \\
   &  &   + g_4^{(3)} g_2^{(1)} - g_4^{(3)} g_1^{(2)}
                   - g_4^{(1)} g_1^{(3)}         \nonumber      \\
   &  &   +  12 \tilde{g}_4^{(1)}   \tilde{g}_2^{(3)}
            + 12 \tilde{g}_4^{(3)}   \tilde{g}_2^{(1)}     )    \\
\frac{ \partial g_4^{(1)}}{\partial l}  & = &
       -  \frac{1}{\pi v_F}  (  g_4^{(1)} g_4^{(1)}
     +  g_2^{(1)} g_2^{(1)}  -  g_1^{(2)} g_2^{(1)}   \nonumber  \\
   &  &  +  g_2^{(3)} g_2^{(3)}  -  g_2^{(3)} g_1^{(3)}   \nonumber  \\     
   &  &       + 6 \tilde{g}_4^{(1)}   \tilde{g}_4^{(1)}
      + 6 \tilde{g}_2^{(1)}   \tilde{g}_2^{(1)}     \nonumber    \\                                               
    &  &     + 6 \tilde{g}_4^{(3)}   \tilde{g}_4^{(3)}
            + 6 \tilde{g}_2^{(3)}   \tilde{g}_2^{(3)}  )       \\
\frac{ \partial g_4^{(2)}}{\partial l}  & = &
       -  \frac{1}{2 \pi v_F}  (  g_4^{(1)} g_4^{(1)}
           -  g_1^{(2)} g_1^{(2)}                   \nonumber    \\     
   &  &  -  g_1^{(3)} g_1^{(3)}  -  g_4^{(3)} g_4^{(3)}  )      \\
\frac{\partial g_4^{(3)}}{\partial l}  & = &
      (2 - K_{+} - \frac{1}{K_{-}})  g_4^{(3)}
   -  \frac{1}{\pi v_F}   (  g_4^{(3)} g_4^{(1)}   \nonumber   \\
    &  &   + 2  g_2^{(3)} g_2^{(1)} - g_1^{(3)} g_2^{(1)}
                                      \nonumber        \\
   &  &   - g_2^{(3)} g_1^{(2)} - g_1^{(3)} g_1^{(2)}  \nonumber  \\
  &  &   +  12 \tilde{g}_4^{(1)}   \tilde{g}_4^{(3)}
        +  12 \tilde{g}_2^{(1)}   \tilde{g}_2^{(3)}    )
\label{last}
\end{eqnarray}
The variable $l$ stands for minus the logarithm of the energy
(temperature) scale measured in units of the high-energy cutoff
of the 1D model $E_c \sim v_F k_c$ (of the order of $\sim 0.1 \; 
{\rm eV}$).
The large coefficients in front of the intertube contributions
arise from the number of nearest-neighbors of each nanotube in
the 3D array.
Furthermore, we have also incorporated a nonperturbative
improvement of the equations by writing the exact dependence of
the scaling dimensions on the forward-scattering couplings,
expressed in terms of the $K_+, K_-$ parameters.

\begin{figure}
\includegraphics[width =7.5cm ]{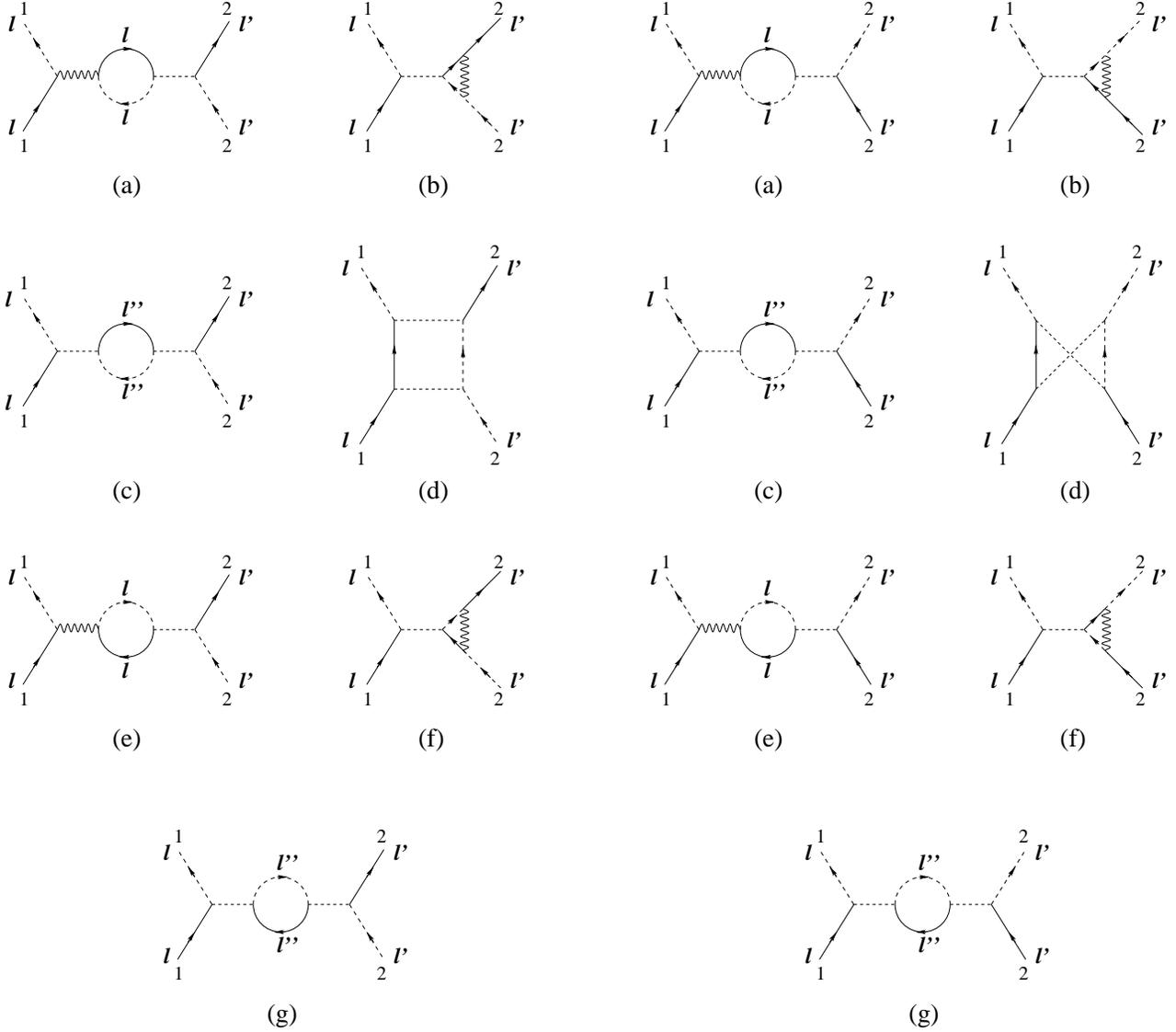}
\caption{Second-order diagrams with logarithmic dependence on the
frequency renormalizing the intertube $\tilde{g}_2^{(1)}$ interaction.
The wavy lines stand for intratube interactions and the dashed lines
(without arrow) for interactions between nearest-neighbor nanotubes
${\bf l, l'}$ in a 3D array. The rest of the lines keep the same
meaning as in Fig. \ref{gtilde}.}
\label{backs}
\end{figure}

\begin{figure}
\includegraphics[width =7.5cm ]{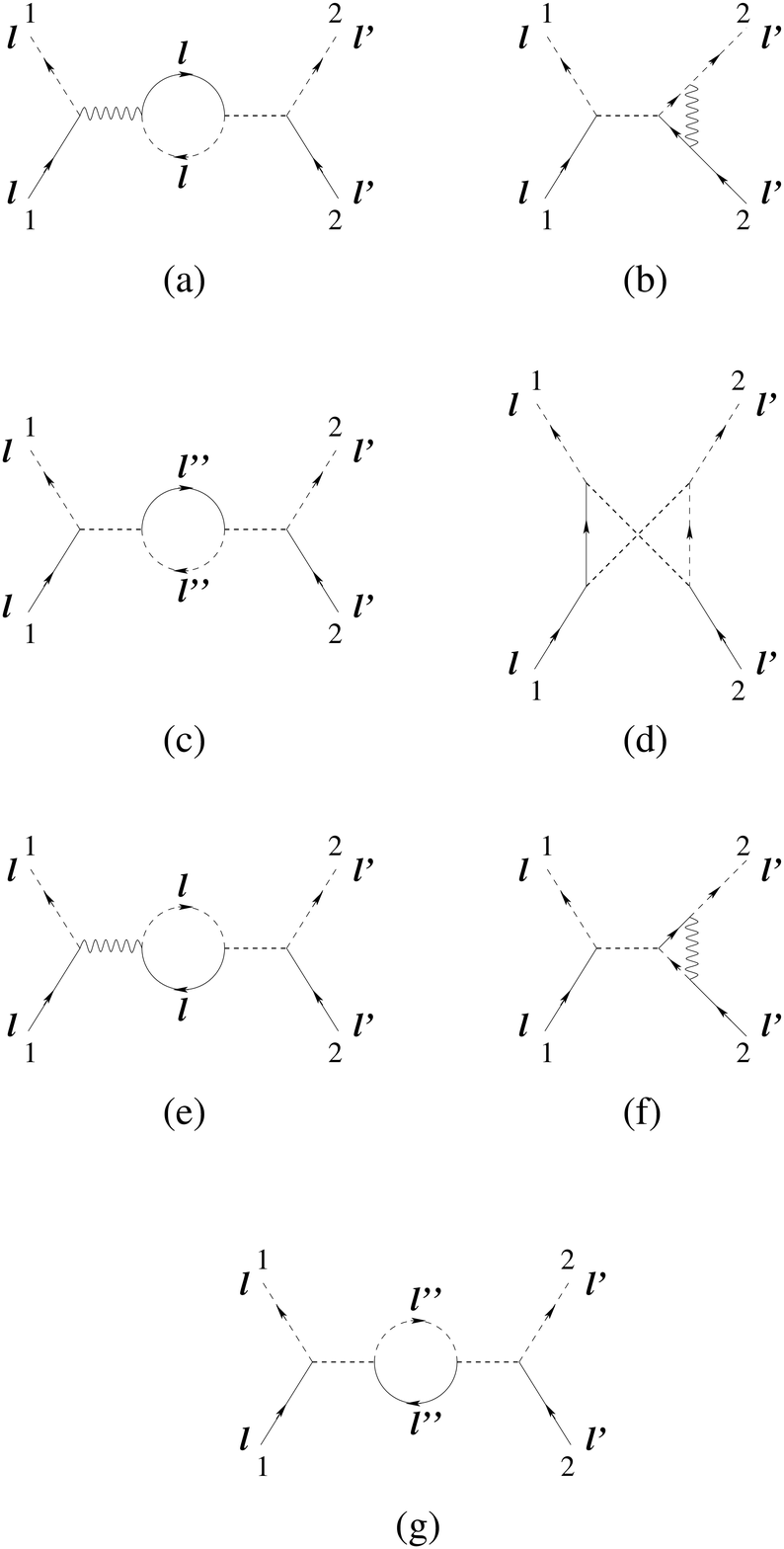}
\caption{Second-order diagrams with logarithmic dependence on the
frequency renormalizing the intertube $\tilde{g}_2^{(3)}$ interaction.
The wavy lines stand for intratube interactions and the dashed lines
(without arrow) for interactions between nearest-neighbor nanotubes
${\bf l, l'}$ in a 3D array.}
\label{umkl}
\end{figure}

The new intertube interactions are themselves corrected by 
processes that diverge logarithmically at low energy, and that
give rise to respective contributions to the scaling of the 
intertube couplings. We can focus on the analysis of 
$\tilde{g}_2^{(1)}, \tilde{g}_4^{(1)}, \tilde{g}_2^{(3)}$
and $\tilde{g}_4^{(3)}$, which have values given by the 
intertube potential (\ref{intert}) at the initial stage of the 
low-energy scaling. As long as this potential decays exponentially 
for $|{\bf l}-{\bf l'}| k > 1$, we can neglect the influence of
other intertube interactions with $2 k_F$ momentum-transfer.
The second-order diagrams that renormalize the above intertube
couplings consist of particle-particle processes or particle-hole
loops involving a change of chirality, as illustrated in 
Figs. \ref{backs} and \ref{umkl}. Some of the contributions
depend on $\tilde{g}_2^{(2)}$ and $\tilde{g}_4^{(2)}$. This means 
that these couplings have to be taken into account for a consistent 
description of the low-energy scaling. The complete set of scaling 
equations for the intertube couplings becomes:
\begin{eqnarray}
\frac{ \partial \tilde{g}_2^{(1)}}{\partial l}  & = &
 - \frac{1}{\pi v_F}  ( 2 g_4^{(1)} \tilde{g}_2^{(1)}
          +  2 g_2^{(1)} \tilde{g}_4^{(1)}
       + 4  \tilde{g}_4^{(1)} \tilde{g}_2^{(1)}    \nonumber     \\
   &   &  +   \tilde{g}_2^{(2)}  \tilde{g}_2^{(1)}
          -   g_4^{(2)}  \tilde{g}_2^{(1)}
          -   g_1^{(2)}  \tilde{g}_4^{(1)}         \nonumber      \\
   &   &  +  2 g_2^{(3)}  \tilde{g}_4^{(3)}
          +   g_4^{(3)}  \tilde{g}_2^{(3)}
          +  4 \tilde{g}_2^{(3)} \tilde{g}_4^{(3)}  \nonumber      \\
   &   &  -   g_1^{(3)}  \tilde{g}_4^{(3)}     )                 \\
\frac{ \partial \tilde{g}_2^{(3)}}{\partial l}  & = &
 - \frac{1}{\pi v_F}  ( 2 g_4^{(1)} \tilde{g}_2^{(3)}
          +  2 g_2^{(3)} \tilde{g}_4^{(1)}
       + 4  \tilde{g}_4^{(1)} \tilde{g}_2^{(3)}    \nonumber     \\
   &   &  +   g_4^{(3)}  \tilde{g}_2^{(1)}
          +  2 g_2^{(1)}  \tilde{g}_4^{(3)}
          +  4 \tilde{g}_2^{(1)}  \tilde{g}_4^{(3)} \nonumber      \\
   &   &  -   g_1^{(3)}  \tilde{g}_4^{(1)}
          -   g_4^{(2)}  \tilde{g}_2^{(3)}
          -   g_1^{(2)} \tilde{g}_4^{(3)}          \nonumber      \\
   &   &  -   \tilde{g}_2^{(2)}  \tilde{g}_2^{(3)}   )        \\
\frac{ \partial \tilde{g}_4^{(1)}}{\partial l}  & = &
 - \frac{1}{\pi v_F}  ( 2 g_4^{(1)} \tilde{g}_4^{(1)}
          +  2 g_2^{(1)} \tilde{g}_2^{(1)}          \nonumber     \\
   &   &    + 2  \tilde{g}_4^{(1)} \tilde{g}_4^{(1)}
         + 2  \tilde{g}_2^{(1)} \tilde{g}_2^{(1)}
          +   \tilde{g}_4^{(2)}  \tilde{g}_4^{(1)}    \nonumber    \\
   &   &   -   g_4^{(2)}  \tilde{g}_4^{(1)}
           -   g_1^{(2)}  \tilde{g}_2^{(1)}         \nonumber      \\
   &   &    + 2  g_2^{(3)} \tilde{g}_2^{(3)}
         +  2 \tilde{g}_2^{(3)} \tilde{g}_2^{(3)}
          +   g_4^{(3)}  \tilde{g}_4^{(3)}    \nonumber    \\
   &   &    +  2 \tilde{g}_4^{(3)} \tilde{g}_4^{(3)}
            -   g_1^{(3)} \tilde{g}_2^{(3)}    )                 \\
\frac{ \partial \tilde{g}_4^{(3)}}{\partial l}  & = &
 - \frac{1}{\pi v_F}  ( 2 g_4^{(1)} \tilde{g}_4^{(3)}
          +   g_4^{(3)} \tilde{g}_4^{(1)} 
          +  4 \tilde{g}_4^{(1)} \tilde{g}_4^{(3)}     \nonumber     \\
   &   &    + 2  g_2^{(1)} \tilde{g}_2^{(3)}
           +  2  g_2^{(3)} \tilde{g}_2^{(1)}
          +  4 \tilde{g}_2^{(1)}  \tilde{g}_2^{(3)}    \nonumber    \\
   &   &   -   g_4^{(2)}  \tilde{g}_4^{(3)}
           -   \tilde{g}_4^{(2)}  \tilde{g}_4^{(3)}    \nonumber      \\
   &   &    -   g_1^{(2)} \tilde{g}_2^{(3)}
           -   g_1^{(3)} \tilde{g}_2^{(1)}    )                   \\
\frac{ \partial \tilde{g}_4^{(2)}}{\partial l}  & = &
    - \frac{1}{2 \pi v_F} ( \tilde{g}_4^{(1)} \tilde{g}_4^{(1)} 
         -  \tilde{g}_4^{(3)} \tilde{g}_4^{(3)}  )    \label{g42}     \\
\frac{ \partial \tilde{g}_2^{(2)}}{\partial l}  & = &
    - \frac{1}{2 \pi v_F}  ( \tilde{g}_2^{(1)} \tilde{g}_2^{(1)} 
       -  \tilde{g}_2^{(3)} \tilde{g}_2^{(3)}   )     \, .  \label{g22}
\end{eqnarray}

Following the flow of the scaling equations, the backscattering 
and Umklapp interactions are progressively enhanced as the theory is 
scaled down to low energies. At the initial stage of the renormalization, 
the values of the couplings are dictated by the Coulomb interaction and,
in the case of intratube couplings, also by the
effective interaction mediated by phonon-exchange. Regarding the 
Coulomb interaction, its contribution to forward-scattering
couplings is given by the RPA calculation exposed above. Thus we have 
$g_4^{(4)}=g_2^{(4)}=g_2^{(2)}=g_4^{(2)}=V^{(r)}_{{\bf l},{\bf l}}(k=0)/v_F$, 
following the trend shown in Fig. \ref{intra-k}, while 
$\tilde{g}_2^{(2)}=\tilde{g}_4^{(2)}=V^{(r)}_{{\bf l},{\bf l'}}(k=0)/v_F$, 
with the potential for nearest-neighbor ${\bf l},{\bf l'}$ represented 
in Fig. \ref{inter-k}. For the chirality-breaking processes,
the different symmetry of ingoing and outgoing electron modes implies 
also a significant reduction of the Coulomb potential, as evaluated in 
Ref. \onlinecite{eg}. The result is that, for 
the armchair (3,3) nanotubes, there is a repulsive component in the 
backscattering and Umklapp interactions at small momentum-transfer 
whose strength can be estimated as $\approx 0.23 v_F /\kappa $. 
This applies to the couplings 
$g_2^{(1)}, g_2^{(3)}, g_4^{(1)}$ and  $g_4^{(3)}$, as well as
to the corresponding
intertube couplings at small momentum-transfer. For the 
intratube couplings with the large momentum-transfer $2 k_F$, the 
contribution by the Coulomb interaction can be obtained from the
Fourier transform of the potential (\ref{potent}). The strength
estimated in this way turns out to be $\approx 0.06 v_F /\kappa $.

Dealing now with the contribution from the effective phonon-mediated
interaction, we note that the situation is reversed, and that the 
backscattering and Umklapp couplings at vanishing momentum-transfer
get a smaller component than the couplings for momentum-transfer
around $2 k_F$. Previous estimates had already found that the ratio 
between these two different strengths is approximately 1/3 \cite{caron}. 
More recent calculations of the phonon spectrum by means of density 
functional theory have led to a similar proportion\cite{ab1}.
More precisely, it has been found that the contributions of all the 
phonons with momentum-transfer 
$2 k_F$ add to an effective coupling $\lambda \approx 0.1$, while the 
contributions of the phonons near the zone center give an effective
coupling $\approx \lambda/3$. These more accurate estimates are about
three times smaller than those quoted in Ref. \onlinecite{caron}.
Anyhow, we will consider the phase diagram of the (3,3) nanotubes
by spanning a suitable range in the scale $\lambda $
of the two effective couplings, covering the values between the different 
estimates in Refs. \onlinecite{ab1} and \onlinecite{caron}.

\begin{figure}
\includegraphics[width =7.5cm ]{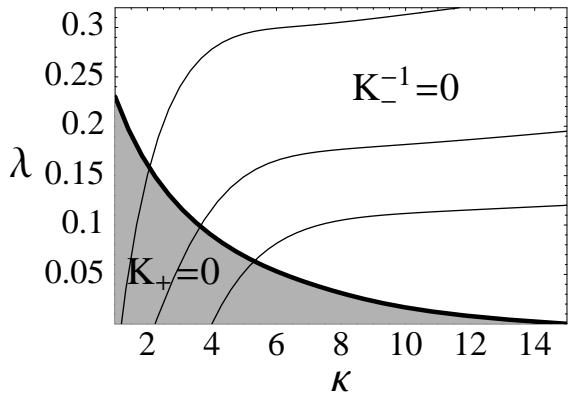}
\caption{Phase diagram showing the different low-energy instabilities
in the array of (3,3) nanotubes, depending on the dielectric constant 
$\kappa $ and the effective coupling $\lambda $ of the phonon-mediated
interaction. The three thin curves correspond to constant $l_{c}$
lines in the phase diagram: $l_{c}=4,6,8$ from top to bottom.}
\label{fase}
\end{figure}

In order to determine the electronic instabilities that may 
appear at low energies, we have solved the set of scaling equations 
(\ref{first})-(\ref{g22}), taking initial values for the couplings 
according to the above discussion. The couplings approach in general
a regime where they grow large as $l \rightarrow \infty $. Regarding
the forward-scattering interactions, $g_4^{(2)}$ becomes increasingly
repulsive, leading to a singularity characterized by either the 
vanishing of $K_+$ or the divergence of $K_-$, depending on the region
of the phase diagram. The scaling of the interactions stops at the 
low-energy scale $\omega_c \equiv E_c e^{-lc}$ corresponding to the
point $l_c$ where the singularity is reached. This has actually the 
character of a critical point, since it gives rise to the opening of
a branch cut and nonanalytic behavior in the corresponding Luttinger
liquid parameter $K_+$ or $K_-$\cite{LET}. We have plotted in Fig. 
\ref{fase} the phase diagram of the (3,3) nanotubes showing the two 
different regions of singular behavior, as a function of the dielectric 
constant $\kappa$ and the effective coupling $\lambda $ of the 
phonon-mediated e-e interaction.

In general, the enhancement of backscattering and Umklapp interactions
upon scaling may lead to a large growth of electron correlations,
pointing at the tendency towards long-range order in the electron
system. We have analyzed this possibility through the computation
of different response functions 
\begin{equation}
\chi(k, \omega)   =   
-i\int_{-\infty}^{+\infty} dt  \int_0 ^{L} dx  e^{i \omega t}
e^{ikx} \langle T O(x, t)  O(0,0)^{\dagger}   \rangle 
\end{equation}
where the pair field $O$ characterizes a particular type of 
ordering. In the system under consideration, the most important 
correlation functions are found to be given by the following 
(Fourier transformed) fields:
\begin{eqnarray}
&& O_{DW,\mu}(k \approx 2k_{F})=\frac{1}{2\sqrt{L}} \sum_{p,\alpha,\beta} \left[
\Psi^{\dagger}_{R1 \alpha}(p-k) \sigma^{\alpha, \beta}_{\mu} \Psi_{L2 \beta} (p) 
\right.
\nonumber \\
&& \quad \quad \quad  +  \left.
\Psi^{\dagger}_{L1 \alpha}(p-k) \sigma^{\alpha, \beta}_{\mu} \Psi_{R2 \beta}(p) 
\right] \,
, \nonumber \\
&& O_{DW',\mu}(k \approx 0)=\frac{1}{2\sqrt{L}} \sum_{p,\alpha,\beta} \left[
\Psi^{\dagger}_{R1 \alpha}(p-k) \sigma^{\alpha, \beta}_{\mu} \Psi_{L1 \beta} (p) 
\right.
\nonumber \\
&& \quad \quad \quad  +  \left.
\Psi^{\dagger}_{R2 \alpha}(p-k) \sigma^{\alpha, \beta}_{\mu} \Psi_{L2 \beta}(p) 
\right] \,
, \nonumber\\
&& O_{SC,\mu}(k \approx 0)=\frac{1}{2\sqrt{L}} \sum_{p,\alpha,\beta} \alpha \left[
\Psi_{R1 \alpha}(-p+k) \sigma^{-\alpha, \beta}_{\mu} \Psi_{L2 \beta} (p) \right. 
\nonumber \\&& \quad \quad \quad +  \left.
\Psi_{R2 \alpha}(-p+k) \sigma^{-\alpha, \beta}_{\mu} \Psi_{L1 \beta}(p) \right] \, ,
\nonumber
\end{eqnarray}
where, for density wave (DW) operators, $\mu=0$ corresponds to a 
charge-density wave (CDW) and $\mu=1,2,3$ to a spin-density wave (SDW);
while, for superconducting (SC) operators, $\mu=0$ stands for singlet 
superconductivity and $\mu=1,2,3$ for triplet superconductivity  
($\sigma^{\alpha, \beta}_{\mu} $ are the Pauli matrices, with 
$\sigma^{\alpha, \beta}_{0}= {\bf 1}_{2 \times 2} $).

\begin{figure}
\includegraphics[width =7.5cm ]{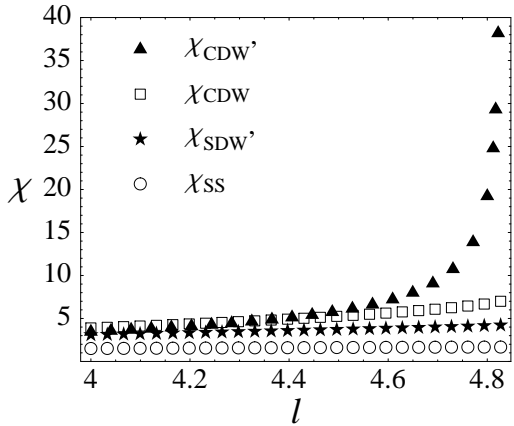}
\caption{Plot of different response functions at $\kappa = 2$ and $\lambda=0.1$.}
\label{Susc}
\end{figure}

The derivatives with respect to the frequency of the response 
functions obey actually scaling equations\cite{sol}, whose solution 
allows to compare the relative growth of the different electron 
correlations. By looking at the low-energy scaling, we have checked 
however that none of the response functions shows a very large growth, 
down to the point where the scaling flow breaks down due to the 
singularity in the Luttinger liquid parameter. This singular behavior 
occurs therefore before the appearance of any tendency to long-range 
order in the electron system. This is illustrated for a typical instance 
in Fig. \ref{Susc}, where it can be observed that only the CDW 
correlations with vanishing momentum show a significant growth as the 
critical point is approached.

The (3,3) carbon nanotubes may fall therefore into 
two different low-temperature phases, whose physical properties are
dictated by the vanishing of $K_{+}$ and the divergence of $K_-$ 
respectively. We remind in particular that, for the experimental setup
of Ref. \onlinecite{tang}, a reasonable choice of the parameters
is $\kappa \sim 2\div 4$ and $\lambda \sim 0.1$, corresponding 
to the $K_+ =0$ phase. In this case the temperature of transition to 
the new phase results strongly dependent on the dielectric constant of 
the environment, ranging from $T_c \sim 10^{-1}$ K (at $\kappa \approx 2 $)
to $T_c \sim 10$ K (at $\kappa \approx 4$).

The behavior of the response functions follows in general the trend 
shown in Fig. \ref{Susc} and, while the density-wave correlations 
tend to grow by approaching the critical value $l_{c}$,
the superconducting correlations remain small anyhow. This finding 
seems to rule out the possibility of having superconducting 
correlations in the (3,3) nanotubes, at least under the physical 
conditions considered in the present paper. We coincide in this respect 
with the conclusions reached in previous analyses by means of other 
computational methods\cite{ab1,ab2}. We have found
however that the appearance of a CDW instability is precluded by the 
breakdown of the Luttinger liquid behavior, which cuts off the growth 
of the different correlations. The destabilization of the Luttinger
liquid is favored in this respect by the screening effects from the 
3D array of nanotubes, which are responsible of bringing the Coulomb 
interaction into the weak-coupling regime.

Our results confront the claim that the experimental signatures
reported in Ref. \onlinecite{tang} should provide evidence for a 
superconducting transition in the small-diameter nanotubes. This 
interpretation has been also challenged by studies of the electron
correlations in the (5,0) nanotubes\cite{ab1,nos}. 
We have shown that, even 
considering the large screening effects from the arrays of nanotubes
in the experimental samples, the effective e-e attraction 
arising from the exchange of phonons is not large enough to support
the appearance of superconducting correlations in the (3,3) 
nanotubes. If the Coulomb interaction is further screened by a 
suitable variation of the dielectric constant of the medium, the 
system is driven then into the phase characterized by the divergence 
of $K_-$ and the related compressibility $\kappa_-$, as shown in 
Fig. \ref{fase}. This has the same character that the instability
given by the Wentzel-Bardeen singularity, where the divergent
compressibility is the signal of the spatial separation of the system
into regions with different electron density\cite{lm,dme}.

Anyhow, the experimental conditions of the samples described in
Ref. \onlinecite{tang} seem to place the system in the region of the
phase diagram characterized by the vanishing of $K_+$. This leads
to the vanishing of the conductivity at the point of the transition,
as follows from Eq. (\ref{drude}). Moreover, it also gives rise to
the vanishing of the tunneling conductance, which is directly 
related to the tunneling density of states $n(\varepsilon )$.
Within the Luttinger liquid framework, the latter follows the 
low-energy behavior
\begin{equation}
n(\varepsilon ) \sim  \varepsilon^{(K_+ + 1/K_+ + K_- + 1/K_- - 4)/8}
\label{dos}
\end{equation}
The depletion of the density of states given by Eq. (\ref{dos})
at vanishing $K_+$ is consistent with the appearance of the pseudogap
observed experimentally in the measures of the $I$-$V$
characteristics reported in Ref. \onlinecite{tang}. The critical point 
characterized by the vanishing of $K_+$ does not describe however a 
conventional metal-insulator transition, as long as the compressibility 
given by Eq. (\ref{compr}) remains finite at the point of the transition. 
As analyzed in Ref. \onlinecite{af}, the critical point implies actually
a divergent diamagnetic susceptibility, as a consequence of the 
development of very soft modes in the sector of electron current
excitations. Therefore, the phenomenology derived from the $K_+ =0$ 
phase of the (3,3) carbon nanotubes seems to be consistent, at least 
qualitatively, with the main experimental signatures reported in 
Ref. \onlinecite{tang}. Further experimental work would be needed to 
clarify the existence of such a critical point in the (3,3) nanotubes,
its physical properties, and its stability under changes of relevant
experimental conditions.

\section*{Acknowledgements}
The financial support of the Ministerio
de Educaci\'on y Ciencia (Spain) through grant
BFM2003-05317 is gratefully acknowledged.
E. P. was also supported by INFN grant 10068.


\end{document}